\documentclass[
prd,superscriptaddress,
twocolumn,
nofootinbib,showpacs
]{revtex4-2}

\usepackage{amsfonts,amssymb,amsmath}
\usepackage{mathrsfs}
\usepackage{xcolor}
\usepackage{graphicx}
\usepackage{dcolumn}
\usepackage{bm}
\usepackage{hyperref}
\usepackage{tabularx}

\begin{document}

\title{Gravitational wave signatures of preheating in Higgs--$R^2$ inflation}

\author{Jinsu Kim}
\email{kimjinsu@tongji.edu.cn}
\affiliation{
    School of Physics Science and Engineering, 
    Tongji University, 
    Shanghai 200092, China
}
\author{Zihao Yang}
\email{2230985@tongji.edu.cn}
\thanks{Co-first Author}
\thanks{Corresponding Author}
\affiliation{
    School of Physics Science and Engineering, 
    Tongji University, 
    Shanghai 200092, China
}
\author{Ying-li Zhang}
\email{yingli@tongji.edu.cn}
\thanks{Co-corresponding Author}
\affiliation{
    School of Physics Science and Engineering, 
    Tongji University, 
    Shanghai 200092, China
}
\affiliation{
    Institute for Advanced Study of Tongji University,
    Shanghai 200092, China
}
\affiliation{
    Institute of Theoretical Physics,
    Chinese Academy of Sciences,
    Beijing 100190, China
}
\affiliation{
    Kavli Institute for the Physics and Mathematics of the Universe (WPI),
    The University of Tokyo Institutes for Advanced Study,
    The University of Tokyo, 
    Chiba 277-8583, Japan
}
\affiliation{
    Center for Gravitation and Cosmology,
    Yangzhou University, 
    Yangzhou 225009, China
}
\date{\today}

\begin{abstract}
We present a comprehensive analysis of the preheating dynamics and associated gravitational wave signatures in the Higgs--$R^2$ inflationary model. Using lattice simulations, we investigate the post-inflationary evolution of the system across the parameter space, covering both the Higgs-like and $R^2$-like scenarios. We demonstrate that the efficiency of preheating is significantly dependent on the nonminimal coupling parameter $\xi$. As the $\xi$ parameter increases, moving from the $R^2$-like regime to the Higgs-like regime, we observe more efficient preheating. Through detailed numerical computations, efficient preheating is shown to lead to stronger gravitational wave production. The amplitude of the gravitational wave spectrum varies by several orders of magnitude as we move from the $R^2$-like regime to the Higgs-like regime. The resultant gravitational wave signatures can serve as a potential observational probe to distinguish between different parameter regimes of the Higgs--$R^2$ model.
\end{abstract}

\maketitle


\section{Introduction}
\label{sec:intro}
The idea of cosmic inflation~\cite{Brout:1977ix,Guth:1980zm,Starobinsky:1980te,Linde:1981mu,Albrecht:1982wi,Linde:1983gd}, originally proposed to resolve several cosmological puzzles, has become an integral part of modern cosmology. Since the birth of the framework, numerous different inflationary models have been proposed over the decades~\cite{Martin:2013tda}. However, recent high-precision measurements of cosmological parameters have significantly constrained viable models. In particular, observations of primordial perturbations imprinted in the cosmic microwave background (CMB) have provided stringent constraints on the scalar spectral index $n_s$ and tensor-to-scalar ratio $r$, dramatically narrowing successful candidates of models~\cite{Planck:2018jri}. 

Nevertheless, some models persist. For example, the $R^2$ model, also known as the Starobinsky model~\cite{Starobinsky:1980te}, and the nonminimal coupling model, commonly referred to as Higgs inflation~\cite{Futamase:1987ua,Cervantes-Cota:1995ehs,Bezrukov:2007ep}, have emerged as remarkably successful. Not only are they theoretically elegant, but they are also the two most favored models by the latest observation~\cite{Planck:2018jri,BICEP:2021xfz}.
The Starobinsky model, one of the earliest proposed inflationary scenarios, introduces a quadratic curvature term $R^2$ to the Einstein-Hilbert action. The model can be recast through the introduction of an auxiliary field, dubbed the scalaron, that drives inflation. Higgs inflation, meanwhile, adopts a scalar field that is nonminimally coupled to gravity through a term $\phi^2 R$. As the Higgs field is the only scalar field discovered so far, the economical idea of using the same scalar field as the inflaton has gained much attention. Despite the different theoretical foundations, both the Starobinsky model and Higgs inflation predict nearly identical values for the scalar spectral index and tensor-to-scalar ratio, namely $n_s \approx 0.965$ and $r \approx 0.003$, which are strongly supported by CMB observations ~\cite{Planck:2018jri,BICEP:2021xfz}.

The success and similarities of the two models naturally motivate their unification. Since both models modify the gravitational sector through mass dimension-4 operators, it is compelling to consider a unified two-field model incorporating both mechanisms. Moreover, the introduction of the scalaron can naturally address the unitarity issue present in Higgs inflation~\cite{Salvio:2015kka,Calmet:2016fsr,Ema:2017rqn}. The unified two-field model is referred to as the Higgs--$R^2$ model, the Starobinsky--Higgs model, or the Higgs--scalaron model, and is widely applied in modern cosmological research; see, for instance, Refs.~\cite{Salvio:2015kka,Kaneda:2015jma,Calmet:2016fsr,Ema:2017rqn,Wang:2017fuy,Pi:2017gih,Wang:2018kly,Gundhi:2018wyz,Gorbunov:2018llf,He:2018gyf,Karam:2018mft,Ghilencea:2018rqg,Cheong:2019vzl,Canko:2019mud,Gundhi:2020zvb,Ema:2021xhq,Cheong:2022gfc,Wang:2024vfv,Pineda:2024prs,Kuralkar:2025hoz,Kim:2025dyi}.
Constraints from CMB observations and theoretical consistency reduce the free parameters of the model to one. Depending on the parameter regime, the model exhibits either Higgs-like or $R^2$-like inflationary behavior. Importantly, these constraints demonstrate that pure Higgs inflation cannot be realized within the system~\cite{Ema:2017rqn}, indicating the importance of the inclusion of the $R^2$ term.

A crucial aspect of any successful inflationary model is the ability to transition from the inflationary phase to the Hot Big Bang Universe through reheating. During inflation, the Universe undergoes exponential expansion driven by the potential energy of the inflaton fields. As inflation ends, these fields begin to oscillate around their potential minimum, and the Universe will witness a transition to a radiation-dominated era. During this reheating phase, the inflaton fields transfer their energy into elementary particles that eventually thermalize and establish the Hot Big Bang Universe. 
The initial stage of the reheating process, known as preheating, proceeds through nonperturbative effects, far from thermal equilibrium. During this phase, two main mechanisms may drive efficient particle production, namely parametric resonance and tachyonic instability. Parametric resonance occurs when the periodic oscillations of the inflaton fields lead to resonant amplification of specific momentum modes through the periodic variation of their effective mass. On the other hand, tachyonic instability arises when the effective mass-squared of fluctuations becomes temporarily negative during the oscillations, leading to exponential growth of certain modes.
For a review on the topic, readers may refer to Refs.~\cite{Amin:2014eta,Lozanov:2020zmy}.
Earlier studies on the preheating phase for the Higgs--$R^2$ model have shown that both parametric resonance and tachyonic instability mechanisms play important roles in the dynamics; see, {\it e.g.}, Refs.~\cite{Bezrukov:2019ylq,Bezrukov:2020txg,He:2020ivk}. The importance and interplay of these effects depend on the model parameters, particularly the ratio of the nonminimal coupling and the coefficient of the $R^2$ term.

The amplification of the field inhomogeneities during preheating may serve as an efficient source of gravitational wave (GW) production. It has been shown that the nonlinear dynamics during the preheating phase can generate a significant stochastic GW background \cite{Khlebnikov:1997di,Bassett:1999mt,Easther:2006gt,Easther:2006vd,Dufaux:2007pt,Garcia-Bellido:2007fiu,Figueroa:2011ye,Figueroa:2016ojl,Figueroa:2017vfa,Tranberg:2017lrx,Krajewski:2022ezo,Adshead:2024ykw}. During this process, the violent oscillations of the inflaton fields and the tachyonic instability create a highly inhomogeneous energy distribution, sourcing the production of GWs. The characteristic frequency and amplitude of these GWs are determined by the dynamics of the preheating process, which, in turn, depends on the underlying inflationary model and its parameters. In multi-field models like the Higgs--$R^2$ model, the presence of multiple dynamical fields and their interactions can lead to significant GWs. 
The resulting GW spectrum carries important information about both the preheating process and the underlying model parameters, potentially providing a unique observational window into the physics of the early Universe. While the typical frequencies of GWs from preheating are generally too high for current detectors, they motivate the importance and the necessity of novel high-frequency GW experiments and could become accessible to future high-frequency GW experiments.

In this work, we perform a comprehensive analysis of the preheating dynamics and the associated GW signatures in the Higgs--$R^2$ model. We select seven benchmark points (BPs) spanning the parameter space from the $R^2$-like scenario to the Higgs-like scenario, including the mixed scenario. Performing lattice simulations, we track the evolution of field components and energy densities during preheating after the end of inflation, focusing particularly on the enhancement of inhomogeneous modes. We numerically compute the GW signatures produced during the preheating phase, providing predictions for their frequency distribution and amplitude.

This paper is organized as follows. In Sec.~\ref{sec:model}, we present the Higgs--$R^2$ model and summarize the relevant equations. We then analyze in Sec.~\ref{sec:inflation} the inflationary dynamics, adopting the standard slow-roll approximation. The post-inflationary phase, with a focus on preheating, is discussed in Sec.~\ref{sec:preheating}, where we employ lattice simulations to study the development of field inhomogeneities and compute the resulting GW spectrum. Finally, we conclude in Sec.~\ref{sec:conclusion}.

\section{Model}
\label{sec:model}
We consider the Higgs--$R^2$ inflationary model which is described by the action
\begin{align}
  S &= \int d^4x \, \sqrt{-g_{\rm J}} \, \bigg[
  \frac{M^2}{2}R_{\rm J} + \alpha R_{\rm J}^2
  + \frac{1}{2}\xi \phi^2 R_{\rm J}
  \nonumber\\
  &\qquad\qquad\qquad\qquad
  - \frac{1}{2}g^{\mu\nu}_{\rm J}\partial_\mu \phi \partial_\nu \phi
  - V_{\rm J}(\phi)
  \bigg]\,,
  \label{eqn:original-action}
\end{align}
where $M^2 \equiv M_{\rm P}^2 - \xi v^2$, with $M_{\rm P}$ being the reduced Planck mass and $v$ the vacuum expectation value of the $\phi$ field, the subscript J denotes that the action is given in the Jordan frame, and $V_{\rm J}(\phi)$ is the scalar potential of the $\phi$ field.
A scalar-tensor theory that is mathematically equivalent to the action \eqref{eqn:original-action} can be obtained by introducing an auxiliary field $\psi$~\cite{DeFelice:2010aj}:
\begin{align}
    S &= \int d^4x \, \sqrt{-g_{\rm J}} \, \bigg[
    \frac{M^2}{2}\left(
    1+4\alpha\frac{\psi}{M^2}+\xi\frac{\phi^2}{M^2}
    \right)R_{\rm J}
    \nonumber\\
    &\qquad\qquad\qquad\quad
    -\frac{1}{2}g^{\mu\nu}_{\rm J}\partial_\mu\phi\partial_\nu\phi
    -\alpha \psi^2
    -V_{\rm J}(\phi)
    \bigg]
    \,.
    \label{eqn:jordan-action}
\end{align}
We can easily verify that varying the action \eqref{eqn:jordan-action} with respect to the $\psi$ field results in $\psi=R_{\rm J}$ from which the original action \eqref{eqn:original-action} can be recovered. The auxiliary field $\psi$ is commonly referred to as the scalaron whose mass dimension is two, and we refer to the $\phi$ field as the Higgs field.\footnote{
We stress that the $\phi$ field is not the Standard Model Higgs field. Studies of preheating for the Standard Model Higgs inflation model include Refs.~\cite{Bezrukov:2008ut,Garcia-Bellido:2008ycs,Repond:2016sol,Sfakianakis:2018lzf,Hamada:2020kuy}.
}
In this work, we consider $V_{\rm J}(\phi) = \lambda \phi^4/4$ together with $v=0$.\footnote{
A nonzero vacuum expectation value $v \neq 0$ may exhibit an interesting preheating phenomenon such as the oscillon formation; see, {\it e.g.}, Ref.~\cite{Kim:2021ipz}.
}
We thus set $M=M_{\rm P}$ from now on.

\begin{figure}[t!]
    \centering
    \includegraphics[width=0.93\linewidth]{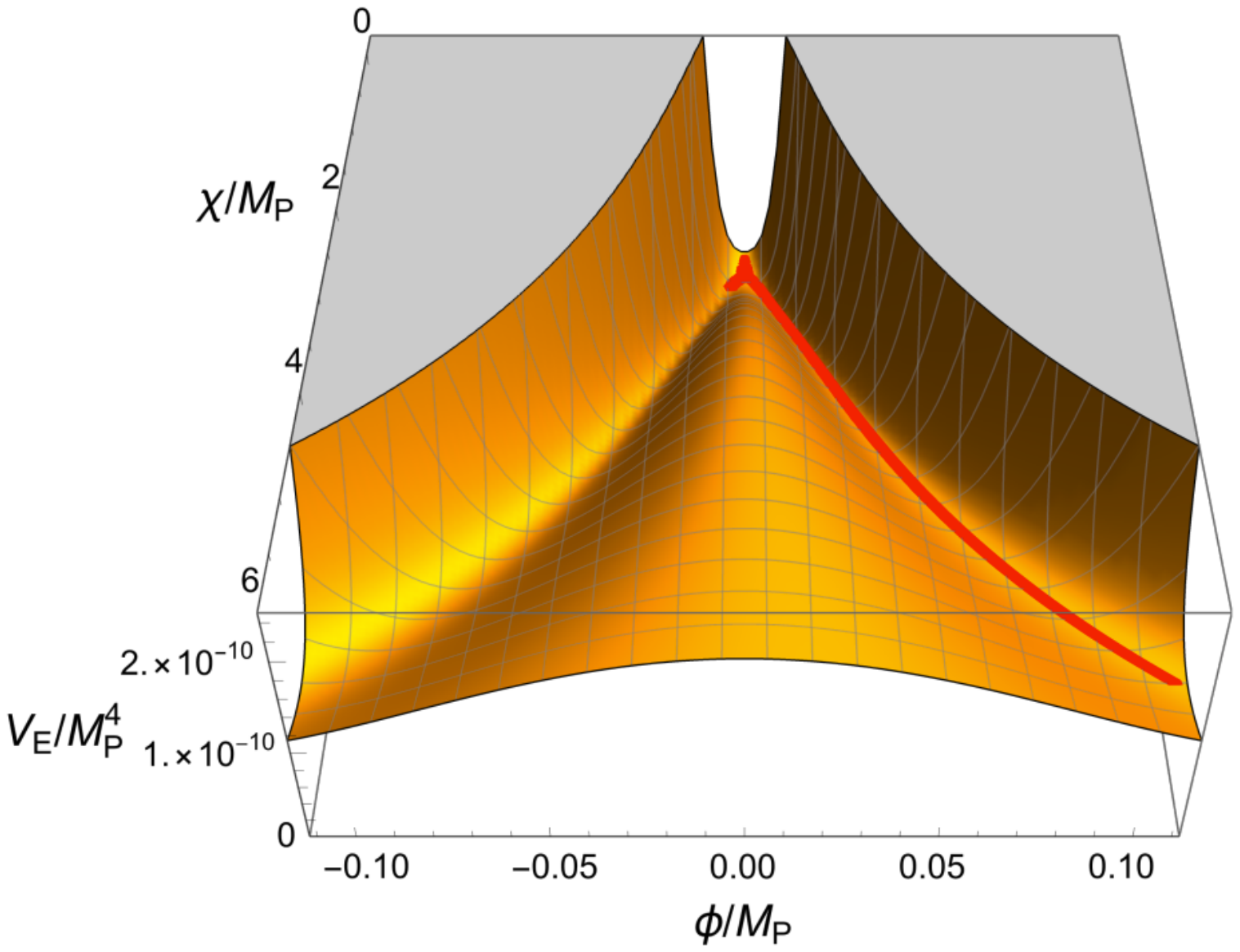}
    \caption{A prototypical Einstein-frame potential. An example of the inflationary trajectory is overlaid in red. The parameters are chosen as $\{\lambda,\xi,\alpha\}=\{0.01, 3464.1, 3.0\times10^8\}$. For the inflationary trajectory, the initial conditions are chosen as $\{\chi/M_{\rm P},\phi/M_{\rm P}\}=\{5.4,0.1\}$, with zero velocities.}
    \label{fig:potential-shape}
\end{figure}

One may bring the Jordan-frame action \eqref{eqn:jordan-action} to the Einstein frame where the gravity sector takes the standard Einstein-Hilbert action via the conformal transformation or Weyl rescaling,
\begin{align}
    g_{{\rm E} \mu\nu} = \Omega^2 g_{{\rm J}\mu\nu}
    \,,
\end{align}
where the conformal factor $\Omega^2$ is given by
\begin{align}
    \Omega^2=1+4\alpha \frac{\psi}{M_{\rm P}^2}+\xi\frac{\phi^2}{M_{\rm P}^2}
    \,,
\end{align}
and the subscript E indicates that it is the Einstein-frame quantity.
Introducing
\begin{align}
    \chi = \frac{\sqrt{6}}{2}M_{\rm P}\ln\left(
    1+4\alpha \frac{\psi}{M_{\rm P}^2}+\xi\frac{\phi^2}{M_{\rm P}^2}
    \right)
    \,,
    \label{eqn:chidef}
\end{align}
the resultant action in the Einstein frame is given by
\begin{align}
  S &= \int d^4x \, \sqrt{-g_{\rm E}} \, \bigg[
  \frac{M_{\rm P}^2}{2}R_{\rm E}
  - \frac{1}{2}g_{\rm E}^{\mu\nu}\partial_\mu \chi \partial_\nu \chi
  \nonumber\\
  &\qquad\qquad\qquad\quad
  - \frac{1}{2}g_{\rm E}^{\mu\nu}f(\chi)\partial_\mu \phi \partial_\nu \phi
  - V_{\rm E}(\phi,\chi)
  \bigg]
  \,,
  \label{eqn:einstein-action}
\end{align}
where the Einstein-frame potential is expressed as
\begin{align}
    V_{\rm E}(\phi,\chi) &= \frac{M_{\rm P}^4}{16\alpha}e^{-2\sqrt{\frac{2}{3}}\frac{\chi}{M_{\rm P}}}
    \nonumber\\&\times
    \left[4\lambda\alpha \frac{\phi^4}{M_{\rm P}^4}+\left(e^{{\sqrt{\frac{2}{3}}\frac{\chi}{M_{\rm P}}}}-1-\frac{\xi \phi^2}{M_{\rm P}^2}\right)^2\right]
    \,,
    \label{eqn:einstein-potential}
\end{align}
and the kinetic coupling function $f(\chi)$ is given by
\begin{align}
    f(\chi) = e^{-\sqrt{\frac{2}{3}}\frac{\chi}{M_{\rm P}}}
    \,.
\end{align}
In Fig.~\ref{fig:potential-shape}, a prototypical shape of the Einstein-frame potential \eqref{eqn:einstein-potential} is shown together with a representative inflationary trajectory.
In the following, as our analysis is done only in the Einstein frame, we omit the subscript E for notational brevity. 

\section{Inflationary Dynamics}
\label{sec:inflation}
Inflationary analysis of the model \eqref{eqn:einstein-action} may proceed by solving the equations of motion. Taking the flat Friedmann-Robertson-Walker (FRW) metric, the homogeneous background fields obey the following equations of motion:
\begin{align}
    \ddot \chi &=
    -V_{,\chi}-3H\dot\chi+\frac{1}{2}f_{,\chi}\dot\phi^2
    \,,\label{eqn:bkg-eom-chi}\\
    \ddot \phi &=
    -f^{-1}V_{,\phi}-3H\dot\phi-f_{,\chi}f^{-1}\dot\chi\dot\phi
    \,,\label{eqn:bkg-eom-phi}
\end{align}
where the dot represents differentiation with respect to the cosmic time $t$, the comma denotes differentiation with respect to the field, and $H$ is the Hubble parameter $H=\dot a/a$, with $a$ being the scale factor. The Hubble parameter follows the first Friedmann equation,
\begin{align}
    H^2 &=
    \frac{1}{3M_{\rm P}^2}\left(
    \frac{\dot \chi^2}{2}+f\frac{\dot \phi^2}{2}+V
    \right)
    \,,\label{eqn:bkg-first-friedmann}
\end{align}
which is the energy conservation equation. Taking the first time derivative of the first Friedmann equation \eqref{eqn:bkg-first-friedmann} and using the background field equations of motion \eqref{eqn:bkg-eom-chi} and \eqref{eqn:bkg-eom-phi}, we have the second Friedmann equation,
\begin{align}
    \dot H &=
    -\frac{1}{2M_{\rm P}^2}\left(
    \dot \chi^2+f\dot\phi^2
    \right)
    \,.\label{eqn:bkg-second-friedmann}
\end{align}
The background inflationary trajectory can then be obtained by solving Eqs.~\eqref{eqn:bkg-eom-chi}, \eqref{eqn:bkg-eom-phi}, and \eqref{eqn:bkg-second-friedmann}. As an example, we present in Fig.~\ref{fig:potential-shape} the inflationary trajectory for the choice of the model parameters $\{\lambda,\xi,\alpha\}=\{0.01, 3464.1, 3.0\times10^8\}$ and the initial conditions for the field values $\{\chi/M_{\rm P},\phi/M_{\rm P}\} = \{5.4,0.1\}$ with $\dot{\phi}=\dot{\chi}=0$.

For an analytical understanding of inflationary dynamics, one may consider a trajectory along the valley of the scalar potential $V$~\cite{Ema:2017rqn,He:2018gyf}. From the first and second derivatives of the potential with respect to the $\phi$ field,
\begin{align}
    V_{,\phi} &= e^{-2\sqrt{\frac{2}{3}}\frac{\chi}{M_{\rm P}}} \phi
    \nonumber\\
    &\quad\times
    \left[
    \left(
    \lambda+\frac{\xi^2}{4\alpha}
    \right)\phi^2 - \frac{\xi M_{\rm P}^2}{4\alpha}\left(
    e^{\sqrt{\frac{2}{3}}\frac{\chi}{M_{\rm P}}} - 1
    \right)
    \right]
    \,,
    \label{eqn:dVdphi}
    \\
    V_{,\phi\phi} &=
    3e^{-2\sqrt{\frac{2}{3}}\frac{\chi}{M_{\rm P}}}
    \nonumber\\
    &\quad\times
    \left[
    \left(\lambda+\frac{\xi^2}{4\alpha}\right)\phi^2
    -\frac{\xi M_{\rm P}^2}{12\alpha}\left(
    e^{\sqrt{\frac{2}{3}}\frac{\chi}{M_{\rm P}}} - 1
    \right)
    \right]
    \,,
    \label{eqn:d2Vdphi2}
\end{align}
we see that the minimum and the maximum of the potential are located at
\begin{align}
    \phi_{\rm min} = \pm M_{\rm P} \sqrt{
    \frac{e^{\sqrt{\frac{2}{3}}\frac{\chi}{M_{\rm P}}} -1}{\xi+4\lambda\alpha/\xi}
    }
    \,,\qquad
    \phi_{\rm max}=0
    \,,
    \label{eqn:phiminmax}
\end{align}
where we have assumed positive model parameters. Discussions on other parameter regimes can be found in Ref.~\cite{Ema:2017rqn}. 
Consequently, the potential exhibits two valleys, located at $\phi_{\rm min}$, separated by a hill at $\phi_{\rm max}$, as one may see in Fig.~\ref{fig:potential-shape}. The inflationary trajectory then proceeds along the valleys,
\begin{align}
    V(\chi,\phi_{\rm min}) =
    \frac{M_{\rm P}^4}{4 \left(4 \alpha + \xi^2/\lambda\right)}
    \left(
    1 - e^{-\sqrt{\frac{2}{3}} \frac{\chi}{M_{\rm P}}}
    \right)^2
    \,.
    \label{eqn:potential-valley}
\end{align}
We note that the inflationary trajectory does not necessarily start at a valley in the initial stage of inflation. However, the trajectory would quickly roll down to a valley as long as the $\phi$ field is much heavier than the Hubble parameter, {\it i.e.},
\begin{align}
    \frac{V_{,\phi\phi}}{H^2}
    =
    6\xi\left(
    4+\frac{\xi^2}{\alpha\lambda}
    \right)\left(
    e^{\sqrt{\frac{2}{3}}\frac{\chi}{M_{\rm P}}}-1
    \right)^{-1}
    \gg
    1
    \,.
\end{align}
From the expression \eqref{eqn:potential-valley}, it is evident that the case of $4\alpha\ll\xi^2/\lambda$ corresponds to the Higgs-like scenario, while the $4\alpha\gg\xi^2/\lambda$ case corresponds to the $R^2$-like scenario. The intermediate regime, $4\alpha\simeq\xi^2/\lambda$, represents the case where both the scalaron and the Higgs field equally contribute; we call this case a mixed scenario, following Ref.~\cite{Ema:2017rqn}.

To compute the inflationary observables, such as the spectral index $n_s$ and the tensor-to-scalar ratio $r$, it is useful to introduce the so-called potential slow-roll parameters,
\begin{align}
    \epsilon_V \equiv
    \frac{M_{\rm P}^2}{2}\left(
    \frac{V_{,\chi}}{V}
    \right)^2
    \,,\quad
    \eta_V \equiv
    M_{\rm P}^2\frac{V_{,\chi\chi}}{V}
    \,.
\end{align}
The spectral index and the tensor-to-scalar ratio are then expressed in terms of $\epsilon_V$ and $\eta_V$ as
\begin{align}
  n_s \simeq 1 - 6\epsilon_V + 2\eta_V
  \,,\quad 
  r \simeq 16\epsilon_V
  \,,
\end{align}
in the slow-roll limit.
To evaluate these quantities at the CMB scale, we take 60 $e$-folds before the end of inflation, marked by the condition $\epsilon_H \equiv -\dot{H}/H^2 \simeq \epsilon_V \simeq 1$ at $\chi=\chi_{\rm end}$ which gives $\chi_{\rm end} \approx 0.94M_{\rm P}$. Since the number of $e$-folds is, under the slow-roll assumption, given by
\begin{align}
    N = \int H \, dt \simeq \frac{1}{M_{\rm P}}\int \frac{1}{\sqrt{2\epsilon_V}}\,d\chi
    \,,
\end{align}
we see that
\begin{align}
    \chi(N_{\rm CMB}) \approx
    \sqrt{\frac{3}{2}}M_{\rm P}\ln\left(\frac{4}{3}N_{\rm CMB}\right)
    \,,
\end{align}
where we have taken $\chi(N_{\rm CMB}) \gg \chi_{\rm end}$. For $N_{\rm CMB} = 60$, we have $\chi(N_{\rm CMB}) \approx 5.37M_{\rm P}$. Thus, the spectral index and the tensor-to-scalar ratio are given by
\begin{align}
    n_s &\simeq 1-\frac{2}{N_{\rm CMB}}-\frac{9}{2N_{\rm CMB}^2}
    \approx 0.9654
    \,,\\
    r &\simeq \frac{12}{N_{\rm CMB}^2}
    \approx 0.003
    \,,
\end{align}
being consistent with the latest CMB observations~\cite{Planck:2018jri,BICEP:2021xfz}.

We note that the spectral index $n_s$ and the tensor-to-scalar ratio $r$ are independent of the model parameters, $\xi$, $\alpha$, and $\lambda$, in the slow-roll limit. They are, however, not all independent free parameters. The magnitude of the curvature power spectrum $A_s$, which is, in the slow-roll limit, given by
\begin{align}
    A_s \simeq \frac{V}{24\pi^2M_{\rm P}^4 \epsilon_V}\,,
\end{align}
should match $A_s \approx 2.1 \times 10^{-9}$ \cite{Planck:2018jri}. This normalization gives a constraint,
\begin{align}
    \frac{\xi^2}{\lambda}+4\alpha \approx 2.4\times10^{9}\label{eqn:cmbnormalization}\,.
\end{align}
Therefore, one of the three model parameters is fixed. Throughout the work, we fix $\lambda=0.01$, and thus, we are left with only one free model parameter. We choose this free parameter to be the nonminimal coupling parameter $\xi$. It is worth noting that the Higgs field $\phi$ acquires the effective quartic coupling of $\lambda+\xi^2/(4\alpha)$ around the vacuum, which should be less than $4\pi$ for the model under consideration to be perturbative~\cite{Ema:2017rqn}, {\it i.e.},
\begin{align}
    \lambda+\frac{\xi^2}{4\alpha}\le 4\pi
    \,,
    \label{eqn:perturbative-constraint}
\end{align}
indicating an upper bound on $\xi$.

\section{Preheating}
\label{sec:preheating}
After inflation ends, the fields start to oscillate around the minimum of the scalar potential. During this phase, one may approximate the scalar potential as
\begin{align}
    V(\chi,\phi) &\approx 
    \frac{\lambda}{4}\phi^4
    +\frac{\xi^2}{16\alpha}\phi^4
    +\frac{M_{\rm P}^2}{24\alpha}\chi^2
    -\frac{\xi M_{\rm P}}{4\sqrt{6}\alpha}\phi^2\chi
    \nonumber\\
    &\quad 
    -\frac{M_{\rm P}}{12\sqrt{6}\alpha}\chi^3
    +\frac{7}{432\alpha}\chi^4
    +\frac{\xi}{8\alpha}\phi^2\chi^2
    \,.
    \label{eqn:potential-approx}
\end{align}
We note that terms proportional to $\chi$ and $\chi^3$, namely the last term in the first line and the first term in the second line, respectively, come with a minus sign. Thus, these terms negatively contribute to the potential in the regime where $\chi > 0$. We also note that the terms in the second line are subdominant compared to the terms in the first line. Approximated analytical treatments of the system with the potential \eqref{eqn:potential-approx} are reported in Ref.~\cite{Bezrukov:2019ylq} (see also Ref.~\cite{Bezrukov:2020txg}) for both homogeneous and inhomogeneous parts of the fields.

The role of the negatively contributing terms in the potential \eqref{eqn:potential-approx} becomes apparent when we consider the equations of motion of the field perturbations $\delta\phi$ and $\delta\chi$. Expanding the fields around their homogeneous background parts, the equations of motion of the field perturbations, in Fourier space, take the form
\begin{align}
	&\ddot{\delta\varphi}_{\bf k}
	+3H\dot{\delta\varphi}_{\bf k}
	+\left(
	\frac{k^2}{a^2}
	+m^2_{\varphi,{\rm eff}}
	\right)\delta\varphi_{\bf k}
	\approx 0\,,
\end{align}
where $\delta\varphi_{\bf k} = \{\delta\phi_{\bf k},\delta\chi_{\bf k}\}$, $k\equiv|{\bf k}|$ is the wavenumber, and we have omitted subleading contributions, including the off-diagonal terms; for the full form, readers may refer to, {\it e.g.}, Refs.~\cite{Lalak:2007vi,Bezrukov:2019ylq,He:2020ivk}. Here, $m^2_{\varphi,{\rm eff}}$ represents the effective mass-squared of the field $\varphi$, which can be extracted by taking derivatives of the potential \eqref{eqn:potential-approx} twice with respect to the field. For the fields $\chi$ and $\phi$, we obtain
\begin{align}
	m^2_{\chi,{\rm eff}} &\approx 
	\frac{M_{\rm P}^2}{12\alpha}
	+\frac{\xi}{4\alpha}\phi^2
	\,,\\
	m^2_{\phi,{\rm eff}} &\approx 
	3\left(
	\lambda + \frac{\xi^2}{4\alpha}
	\right)\phi^2
	-\frac{\xi M_{\rm P}}{2\sqrt{6}\alpha}\chi
	\,,
\end{align}
up to the leading order in the background fields.
Notably, the effective mass-squared of the $\phi$ field may become negative, {\it i.e.}, tachyonic, in the $\chi > 0$ regime. In this case, the $\phi$-field quanta grow sharply. A larger $\xi$ value, that is, a smaller $\alpha$, will bring more severe instability.

Based on the analytical understanding, we perform lattice simulations to investigate the behavior of the field evolution. Using the results of the lattice simulations, we then numerically estimate the GW signatures arising from the preheating of the Higgs--$R^2$ model.

\subsection{Lattice Simulations}
\label{subsec:latsim}
In order to fully capture the post-inflationary dynamics of the Higgs--$R^2$ model, we perform lattice simulations. Lattice simulations for a two-field model with a kinetic coupling function have been investigated in, {\it e.g.}, Refs.~\cite{Krajewski:2018moi,Krajewski:2022ezo,Joana:2022uwc,Adshead:2023nhk,Adshead:2024ykw}. The exact Higgs--$R^2$ model has also been explored in Ref.~\cite{Bezrukov:2020txg}.
Based on the method detailed in Ref.~\cite{Krajewski:2018moi}, which is a modification of the symplectic method used in Ref.~\cite{Sainio:2012mw}, we developed a dedicated code for the Higgs--$R^2$ system. For completeness, we briefly explain the numerical method.
Switching to the conformal time, $d\tau = dt/a$, we discretize the Lagrangian with the grid size of $dx$ in all three spatial dimensions:
\begin{align}
    \mathcal{L} &=
    -3M_{\rm P}^2(a')^2V_3
    +\sum_{\bf x} a^2 \bigg[
    \frac{1}{2}(\chi_{\bf x}')^2
    +\frac{1}{2}f(\chi_{\bf x})(\phi_{\bf x}')^2
    \nonumber\\&\quad
    -\frac{1}{2}\frac{G[\chi_{\bf x}]}{dx^2}
    -\frac{1}{2}f(\chi_{\bf x})\frac{G[\phi_{\bf x}]}{dx^2}
    -a^2V(\chi_{\bf x},\phi_{\bf x})
    \bigg]
    \,,\label{eqn:discrete-Lag}
\end{align}
where the prime denotes differentiation with respect to the conformal time, $V_3$ is the spatial volume, $\phi_{\bf x}$ and $\chi_{\bf x}$ are the field values at ${\bf x}$, and $G$ represents the gradient-squared term, {\it i.e.}, $(\nabla\phi)^2 = G[\phi]/dx^2$.
From the discretized Lagrangian \eqref{eqn:discrete-Lag}, we construct the Hamiltonian, which is given by
\begin{align}
    \mathcal{H} &=
    -\frac{p_a^2}{12M_{\rm P}^2V_3}
    +\sum_{\bf x}\bigg[
    \frac{p_{\chi,{\bf x}}^2}{2a^2}
    +\frac{p_{\phi,{\bf x}}^2}{2a^2f(\chi_{\bf x})}
    \label{eqn:discrete-Hamiltonian}\\&\quad
    +\frac{a^2}{2}\frac{G[\chi_{\bf x}]}{dx^2}
    +\frac{a^2}{2}f(\chi_{\bf x})\frac{G[\phi_{\bf x}]}{dx^2}
    +a^4V(\chi_{\bf x},\phi_{\bf x})
    \bigg]
    \,,\nonumber
\end{align}
where $p_a \equiv \partial\mathcal{L}/\partial a'$, $p_{\phi,{\bf x}} \equiv \partial\mathcal{L}/\partial \phi_{\bf x}'$, and $p_{\chi,{\bf x}} \equiv \partial\mathcal{L}/\partial \chi_{\bf x}'$ are canonical conjugate momenta for the scale factor $a$, the $\phi$ field, and the $\chi$ field, respectively.

Armed with the Hamiltonian \eqref{eqn:discrete-Hamiltonian}, the equations of motion can be obtained by using the Hamilton equations, which take the form of $z'=\{z,\mathcal{H}\}\equiv \mathcal{D}_\mathcal{H}z$, where $z$ is either one of $a$, $\phi$, and $\chi$ or one of the conjugate momenta, with $\{\cdot,\cdot\}$ being the Poisson bracket. The formal solution would then be given by $z(\tau) = e^{\tau\mathcal{D}_\mathcal{H}}z(\tau_0)$, where $\tau_0$ is the initial conformal time. For a sum-separable Hamiltonian $\mathcal{H} = \sum_i\mathcal{H}_i$, such as our case of Eq.~\eqref{eqn:discrete-Hamiltonian}, we need to find a set of operators, $e^{\tau\mathcal{D}_{\mathcal{H}_i}}$, that can approximate $e^{\tau\mathcal{D}_\mathcal{H}}$ up to a desired order \cite{Sainio:2012mw}. Closely following Ref.~\cite{Krajewski:2018moi}, we split the Hamiltonian \eqref{eqn:discrete-Hamiltonian} into $\mathcal{H} = \mathcal{H}_1+\mathcal{H}_2+\mathcal{H}_3+\mathcal{H}_4$, where
\begin{align}
    \mathcal{H}_1 &=
    -\frac{p_a^2}{12M_{\rm P}^2V_3}
    \,,\\
    \mathcal{H}_2 &=
    \sum_{\bf x}\frac{p_{\phi,{\bf x}}^2}{2a^2f(\chi_{\bf x})}
    \,,\quad
    \mathcal{H}_3 =
    \sum_{\bf x}\frac{p_{\chi,{\bf x}}^2}{2a^2}
    \,,\quad\\
    \mathcal{H}_4 &=
    \sum_{\bf x}
    \left[
    \frac{a^2}{2}\frac{G[\chi_{\bf x}]}{dx^2}
    +\frac{a^2}{2}f(\chi_{\bf x})\frac{G[\phi_{\bf x}]}{dx^2}
    +a^4V(\chi_{\bf x},\phi_{\bf x})
    \right]
    \,,
\end{align}
and use the fourth-order integrator, with the time step of $d\tau$, composed of second-order integrators \cite{Sainio:2012mw,YOSHIDA1990262},
\begin{align}
    \Phi^{(4)}(d\tau) &=
    \Phi^{(2)}\left(\frac{d\tau}{2-2^{1/3}}\right)
    \nonumber\\&\quad\times
    \Phi^{(2)}\left(-\frac{2^{1/3}d\tau}{2-2^{1/6}}\right)
    \Phi^{(2)}\left(\frac{d\tau}{2-2^{1/3}}\right)
    \,,
\end{align}
where the second-order symplectic integrator is given by \cite{Sainio:2012mw,Krajewski:2018moi}
\begin{align}
    &\Phi^{(2)}(d\tau) = 
    \Phi_{\mathcal{H}_1}\left(\frac{d\tau}{2}\right)
    \circ
    \Phi_{\mathcal{H}_2}\left(\frac{d\tau}{2}\right)
    \circ
    \Phi_{\mathcal{H}_3}\left(\frac{d\tau}{2}\right)
    \\&\quad
    \circ
    \Phi_{\mathcal{H}_4}\left(d\tau\right)
    \circ
    \Phi_{\mathcal{H}_3}\left(\frac{d\tau}{2}\right)
    \circ
    \Phi_{\mathcal{H}_2}\left(\frac{d\tau}{2}\right)
    \circ
    \Phi_{\mathcal{H}_1}\left(\frac{d\tau}{2}\right)
    \,.\nonumber
\end{align}
Here, $\Phi_{\mathcal{H}_i}$ are the integrators associated with the Hamiltonian $\mathcal{H}_i$ that transform the corresponding dynamical variables. Explicitly, the transformations are given by (see also Refs.~\cite{Sainio:2012mw,Krajewski:2018moi})
\begin{widetext}
\begin{align}
    \Phi_{\mathcal{H}_1}(d\tau) \;&:\;
    \Big(
    a,p_a,\phi_{\bf x},p_{\phi,{\bf x}},\chi_{\bf x},p_{\chi,{\bf x}}
    \Big)
    \to 
    \Big(
    a+\frac{\partial\mathcal{H}_1}{\partial p_a}d\tau,p_a,\phi_{\bf x},p_{\phi,{\bf x}},\chi_{\bf x},p_{\chi,{\bf x}}
    \Big)
    \,,\\
    \Phi_{\mathcal{H}_2}(d\tau) \;&:\;
    \Big(
    a,p_a,\phi_{\bf x},p_{\phi,{\bf x}},\chi_{\bf x},p_{\chi,{\bf x}}
    \Big)
    \to 
    \Big(
    a,p_a-\frac{\partial\mathcal{H}_2}{\partial a}d\tau,\phi_{\bf x}+\frac{\partial\mathcal{H}_2}{\partial p_{\phi,{\bf x}}}d\tau,p_{\phi,{\bf x}},\chi_{\bf x},p_{\chi,{\bf x}}-\frac{\partial\mathcal{H}_2}{\partial\chi_{\bf x}}d\tau
    \Big)
    \,,\\
    \Phi_{\mathcal{H}_3}(d\tau) \;&:\;
    \Big(
    a,p_a,\phi_{\bf x},p_{\phi,{\bf x}},\chi_{\bf x},p_{\chi,{\bf x}}
    \Big)
    \to 
    \Big(
    a,p_a-\frac{\partial\mathcal{H}_3}{\partial a}d\tau,\phi_{\bf x},p_{\phi,{\bf x}},\chi_{\bf x}+\frac{\partial\mathcal{H}_3}{\partial p_{\chi,{\bf x}}}d\tau,p_{\chi,{\bf x}}
    \Big)
    \,,\\
    \Phi_{\mathcal{H}_4}(d\tau) \;&:\;
    \Big(
    a,p_a,\phi_{\bf x},p_{\phi,{\bf x}},\chi_{\bf x},p_{\chi,{\bf x}}
    \Big)
    \to 
    \Big(
    a,p_a-\frac{\partial\mathcal{H}_4}{\partial a}d\tau,\phi_{\bf x},p_{\phi,{\bf x}}-\frac{\partial\mathcal{H}_4}{\partial\phi_{\bf x}}d\tau,\chi_{\bf x},p_{\chi,{\bf x}}-\frac{\partial\mathcal{H}_4}{\partial\chi_{\bf x}}d\tau
    \Big)
    \,.
\end{align}
\end{widetext}
For the gradient-squared term, which can be written as \cite{Frolov:2008hy}
\begin{align}
    G[X_{\bf x}] \equiv \frac{1}{2}\sum_{\bf x'} C_{d({\bf x'})}\left(X_{\bf x'} - X_{\bf x}\right)^2\,,
\end{align}
where the summation is over the neighboring points in the discretized grid space, we take into account 26 neighboring points with the coefficients $c_1 = 7/15$, $c_2 = 1/10$, and $c_3 = 1/30$; this choice corresponds to the isotropic discretization C of Ref.~\cite{Frolov:2008hy}.

\begin{figure}[t!]
    \centering
    \includegraphics[width=0.95\linewidth]{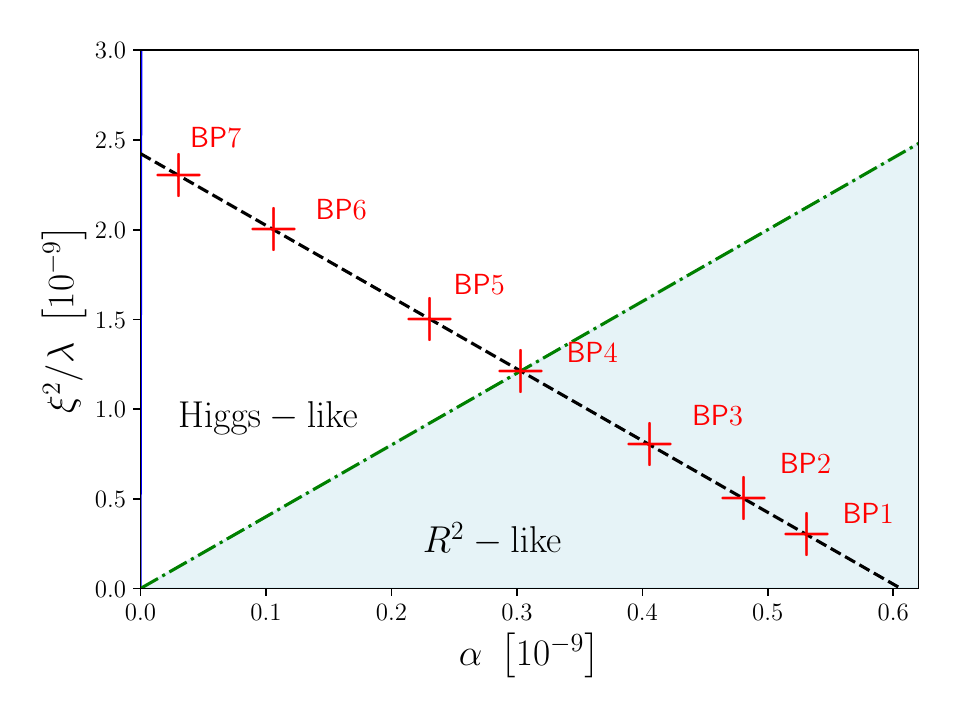}
    \caption{Seven BPs in the $\alpha$--$\xi^2/\lambda$ space. The red cross points depict the BPs outlined in Table~\ref{tab:BPs}. The green dot-dashed line indicates the relation $4\alpha = \xi^2/\lambda$ which separates the $R^2$-like scenario (light green region) and the Higgs-like scenario (white region). All the seven BPs lie on the black dashed line that satisfies the CMB normalization \eqref{eqn:cmbnormalization}. The narrow dark blue region along the $y$-axis represents the nonperturbative region.}
    \label{fig:BPs}
\end{figure}

\begin{table}[t!]
    \renewcommand{\arraystretch}{1.5}  
    \centering
    \begin{tabular}{|c|c|c|c|c|} \hline  
    BP &  $\xi^2/\lambda\left[10^{9}\right]$& $\alpha\left[10^{9}\right]$& $\phi_{\rm ini}\left[10^{-2}M_{\rm P}\right]$& $\phi'_{\rm ini}\left[10^{-8}M^2_{\rm P}\right]$\\ \hline  
    1& 0.3& 0.5312& 1.049& $-2.157$
    \\ \hline  
    2& 0.5& 0.4812& 1.191& $-2.451$
    \\ \hline  
    3& 0.8& 0.4062& 1.340& $-2.757$
    \\ \hline  
    4& 1.2& 0.3032& 1.487& $-3.059$
    \\ \hline  
    5& 1.5& 0.2312& 1.568& $-3.226$
    \\ \hline  
    6& 2.0& 0.1062& 1.685& $-3.467$
    \\ \hline  
    7& 2.3& 0.0312& 1.745& $-3.590$
    \\ \hline
    \end{tabular}
    \caption{Seven BPs. Throughout the simulations, we fix the Higgs quartic coupling to be $\lambda=0.01$. We note that the nonminimal parameter $\xi$ is treated as a free parameter and that $\alpha$ is given from the CMB normalization \eqref{eqn:cmbnormalization} as advertised in Sec.~\ref{sec:inflation}. The initial conditions for the fields and their velocities are given at 0.5 $e$-folds before the end of inflation. The last two columns represent the $\phi$ field values and $\phi'$ values, which depend on the parameter choice. On the other hand, the $\chi$ field and its velocity are independent of the parameter space, and they are given by $\chi_{\rm ini}\approx 1.141M_{\rm P}$ and $\chi'_{\rm ini}\approx-3.055\times10^{-6}M_{\rm P}^2$.}
    \label{tab:BPs}
\end{table}
Throughout the simulations, we use the lattice size of $64^3$. For the grid spacing $dx$, we set it to be $dx = \sqrt{3}\pi/(\Lambda H_{*})$, where $\Lambda$ is the cutoff, and $H_* = M_{\rm P}/\sqrt{12(4\alpha + \xi^2/\lambda)}$ is roughly the Hubble parameter during inflation. In this way, the maximum momentum is given by $k_{\rm max} = \Lambda H_*$. In the current work, we choose $\Lambda = 1000$.
In addition, for all the simulations, we choose a fixed value for the Higgs quartic coupling, $\lambda=0.01$. Then, as advertised in the previous section, the system has only one free parameter, which we choose to be the nonminimal coupling parameter $\xi$. We select seven BPs as shown in Table~\ref{tab:BPs}. The first three BPs correspond to the $R^2$-like scenario, while the last three points correspond to the Higgs-like scenario. The one in the middle, namely BP4, depicts the mixed scenario. These are illustrated in Fig.~\ref{fig:BPs} in the $\alpha$--$\xi^2/\lambda$ parameter space.
The initial conditions for the simulations are determined as follows. We first numerically solve the homogeneous background field equations of motion, Eqs.~\eqref{eqn:bkg-eom-chi}, \eqref{eqn:bkg-eom-phi}, and \eqref{eqn:bkg-second-friedmann}, without any assumptions, for a chosen parameter set. We then identify the end of inflation using the condition $\epsilon_H \equiv -\dot{H}/H^2 = 1$. From this point, we climb 0.5 $e$-folds back and set this point as the initial conditions for the lattice simulation. The initial field values and the velocities are shown in Table~\ref{tab:BPs}.
For the perturbations, we follow the standard Rayleigh distribution, assuming the Gaussian variables; see also Refs.~\cite{Felder:2000hq,Frolov:2008hy}.

\begin{figure*}[ht!]
    \centering
    \includegraphics[width=.5\textwidth]{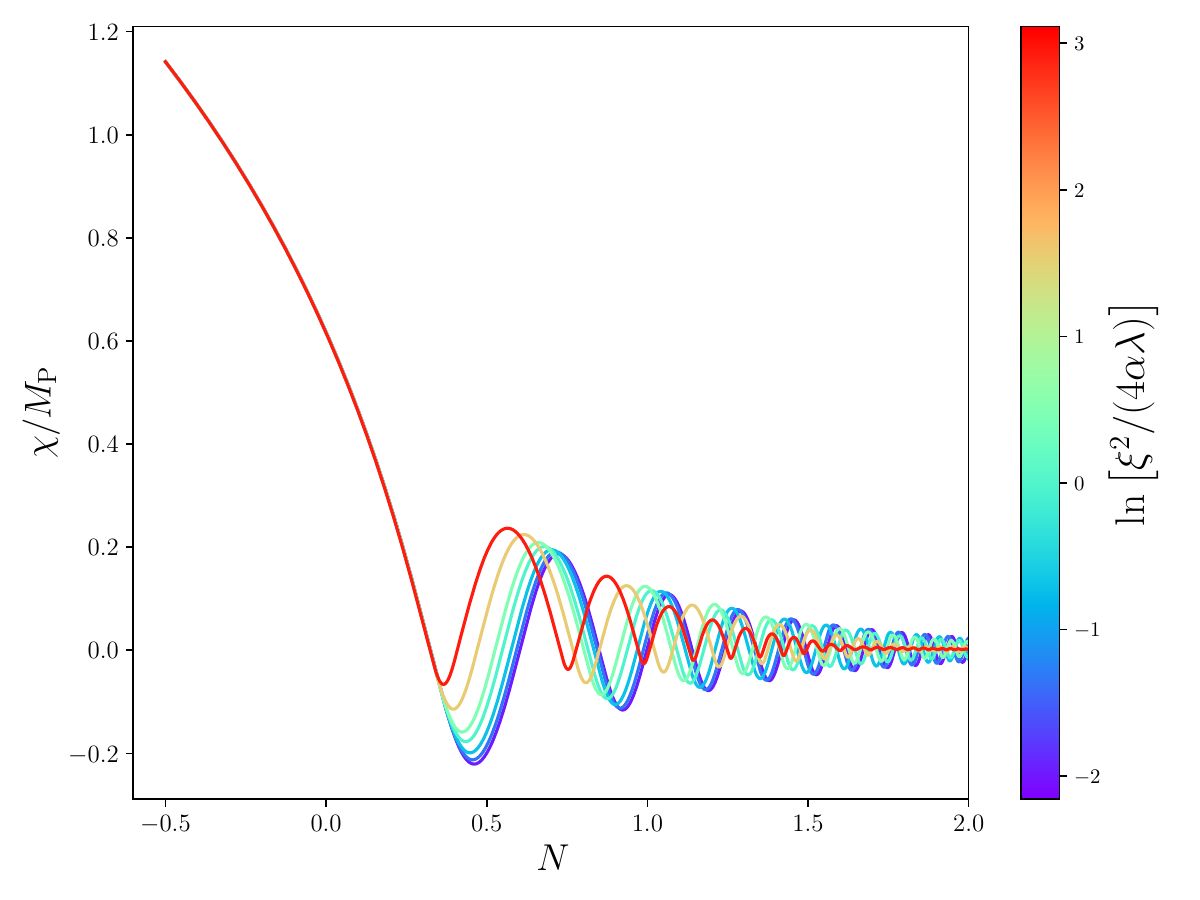}%
    \includegraphics[width=.5\textwidth]{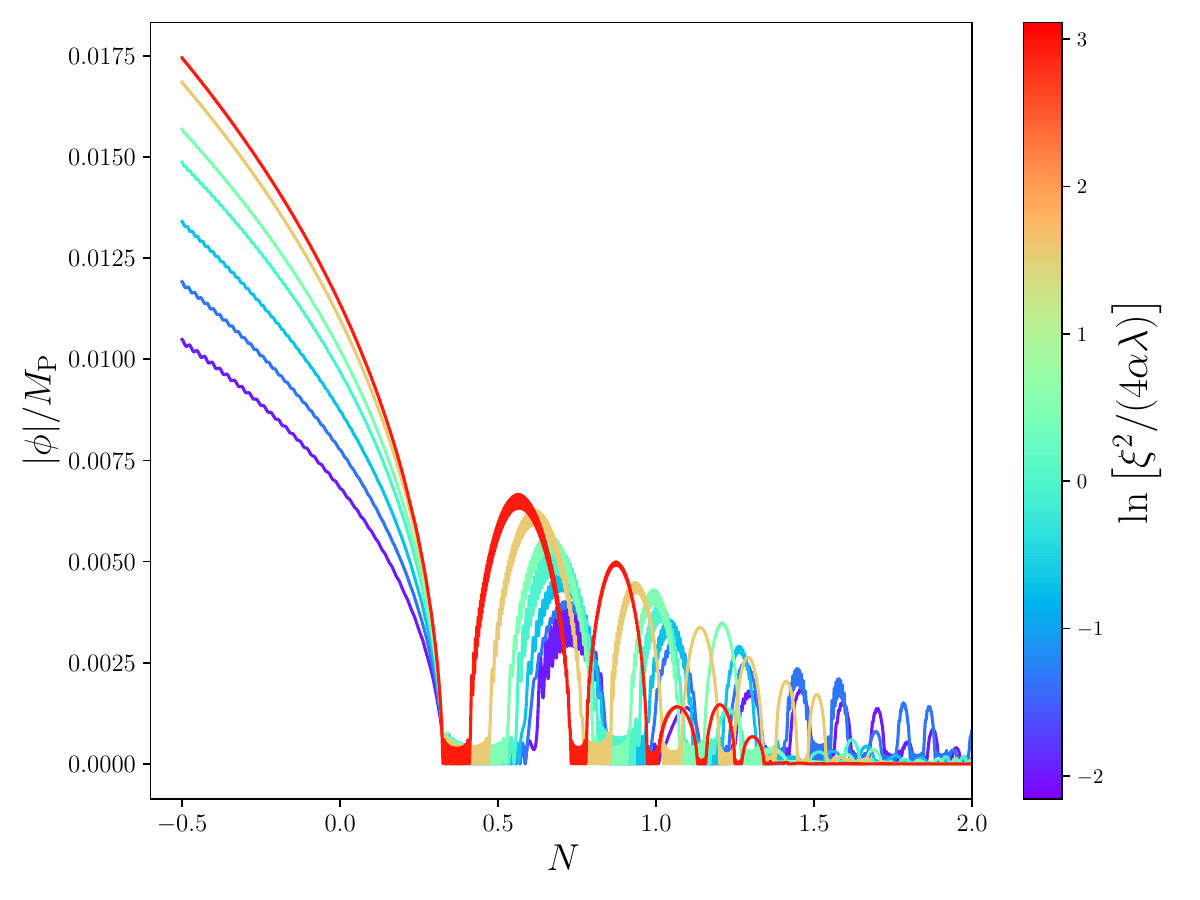}
    \caption{Evolution of the $\chi$ (left) and $\phi$ (right) fields in terms of the number of $e$-folds $N$ for the seven BPs outlined in Table~\ref{tab:BPs}. Different colors indicate different values of the quantity $\ln[\xi^2/(4\alpha\lambda)]$, which takes a negative (positive) value for the $R^2$-like (Higgs-like) scenario. As the nonminimal coupling parameter $\xi$ increases, the oscillation amplitude of the $\chi$ field becomes weaker, while the opposite behavior is observed for the $\phi$ field.}
    \label{fig:fields}
\end{figure*}

Results of the lattice simulations for the seven BPs are as follows.
In Fig.~\ref{fig:fields}, the evolution of the fields is presented. The left panel of Fig.~\ref{fig:fields} shows the evolution of the $\chi$ field in terms of the number of $e$-folds $N$, while the right panel shows the evolution of the absolute value of the $\phi$ field. In both cases, $N=0$ marks the end of inflation, and different colors depict the values of $\ln[\xi^2/(4\alpha\lambda)]$, which is an indicator of whether the scenario is Higgs-like or $R^2$-like; negative (positive) values correspond to the $R^2$-like (Higgs-like) scenario. Before the end of inflation, both the $\chi$ and $\phi$ fields slowly roll down the potential \eqref{eqn:potential-valley}, in good agreement with the discussion given in the previous section. After inflation ends, the fields begin to oscillate around the potential minimum. We observe that the oscillation amplitude of the $\chi$ field becomes larger as the nonminimal coupling parameter $\xi$ decreases. For the $\phi$ field, the opposite tendency is observed.

\begin{figure*}[t!]
    \centering
    \includegraphics[width=.5\textwidth]{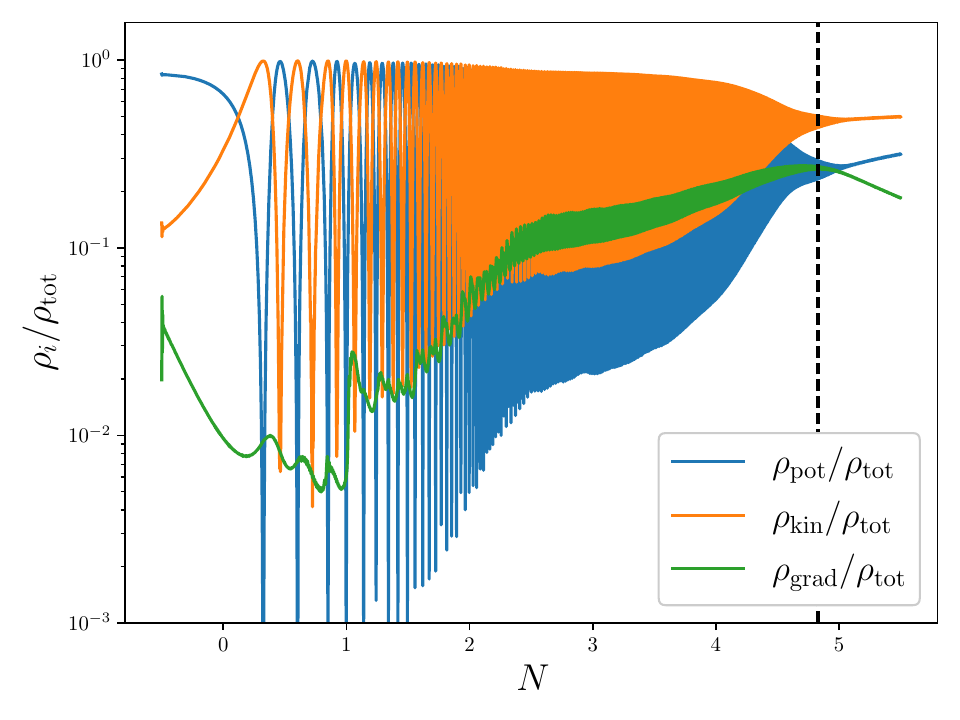}%
    \includegraphics[width=.5\textwidth]{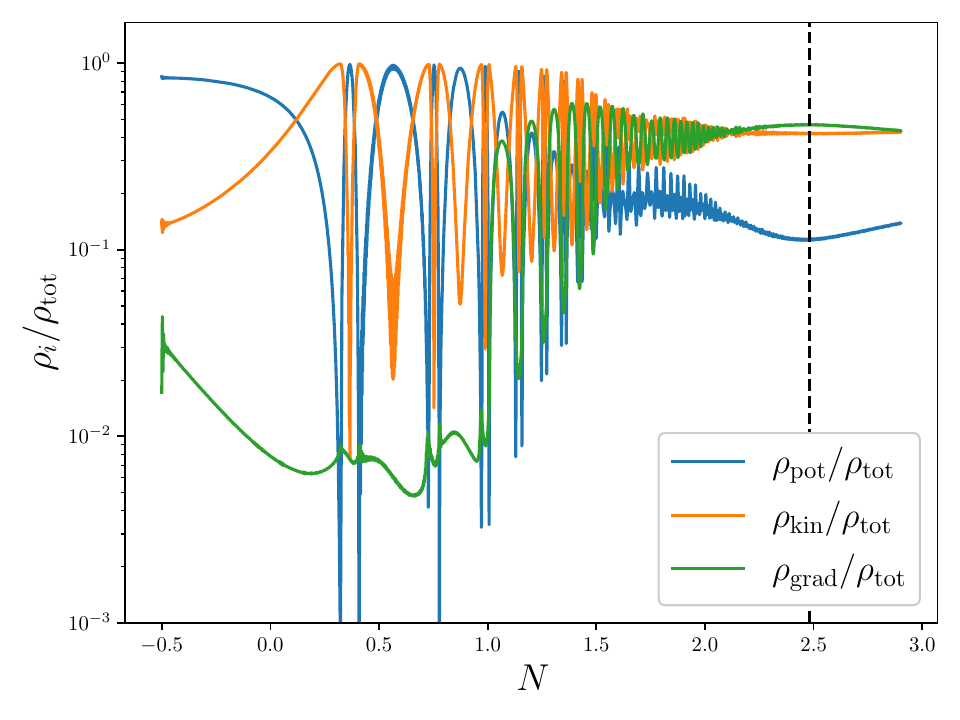}
    \caption{Evolution of the spatially-averaged potential energy density $\rho_{\rm pot}$ (blue line), kinetic energy density $\rho_{\rm kin}$ (orange line), and gradient energy density $\rho_{\rm grad}$ (green line), normalized to the total energy density $\rho_{\rm tot}$ in terms of the number of $e$-folds. The left panel is for BP1, while the right panel is for BP7. One may observe the growth of the inhomogeneity, {\it i.e.}, the gradient energy density, after a couple of oscillations during the post-inflationary regime. The vertical black dashed line in both left and right panels marks the point where the gradient energy density averaged over oscillations reaches its maximum. As the nonminimal coupling parameter $\xi$ increases, moving from the $R^2$-like scenario to the Higgs-like scenario, the growth of the gradient energy density is shown to become stronger and more efficient.}
    \label{fig:energydensity-1-and-7}
\end{figure*}

In Fig.~\ref{fig:energydensity-1-and-7}, the evolutions of the spatially-averaged, normalized kinetic (orange line), gradient (green line), and potential (blue line) energy densities are shown in terms of the number of $e$-folds for BP1 (left panel) and BP7 (right panel). Each energy density component is given by
\begin{align}
    \rho_{\rm pot} &=
    \left\langle V(\chi,\phi) \right\rangle
    \,,\\
    \rho_{\rm kin} &= 
    \left\langle \frac{1}{2} \dot{\chi}^2 + \frac{1}{2}f(\chi)\dot{\phi}^2\right\rangle
    \,,\\
    \rho_{\rm grad} &=
    \left\langle \frac{1}{2a^2}\left(\nabla\chi\right)^2 + \frac{1}{2a^2}f(\chi)\left(\nabla\phi\right)^2 \right\rangle 
    \,,
\end{align}
where $\langle\cdots\rangle$ stands for the spatial average, and the total energy density is $\rho_{\rm tot} = \rho_{\rm pot} + \rho_{\rm kin} + \rho_{\rm grad}$.
As expected, before the end of inflation, the potential energy density slowly decreases, and the kinetic energy slowly increases. After the end of inflation, the two quickly oscillate as the fields undergo oscillations. After a few oscillations, the inhomogeneity rapidly grows, as is manifested in the growth of the gradient energy density. These observations match the approximated analytical understanding sketched in the earlier part of this section as well as in Refs.~\cite{Bezrukov:2019ylq,Bezrukov:2020txg}. In both the left and right panels of Fig.~\ref{fig:energydensity-1-and-7}, the vertical black dashed lines mark the maximum point of the gradient energy density $\rho_{\rm grad}$, averaged over oscillations. They correspond to $N \approx 4.83$ for BP1 and $N \approx 2.48$ for BP7.
We note that the growth of the gradient energy density becomes stronger and more efficient as the nonminimal coupling parameter $\xi$ increases, or, equivalently, as the coefficient of the $R^2$ term $\alpha$ decreases, moving from the $R^2$-like scenario to the Higgs-like scenario.

\begin{figure}[t!]
    \centering
    \includegraphics[width=0.98\linewidth]{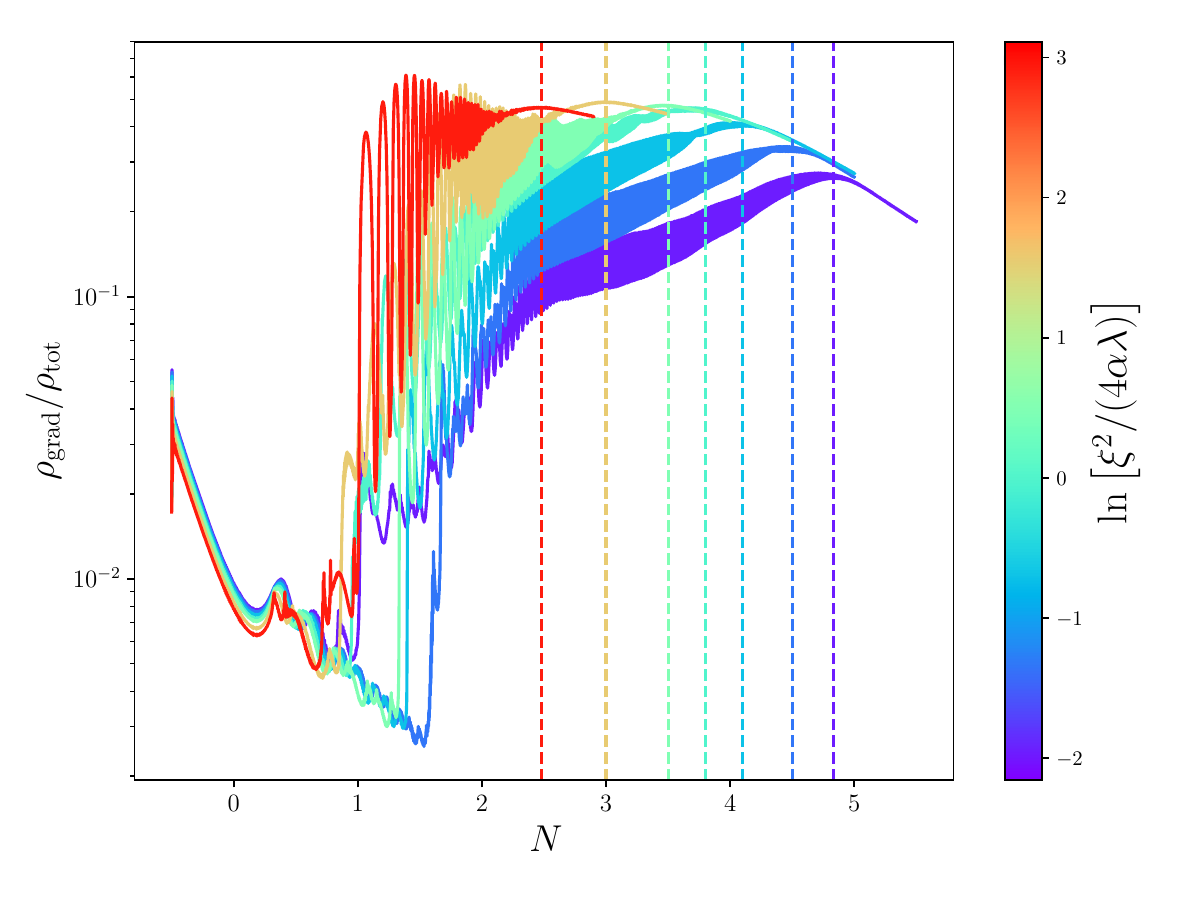}
    \caption{Evolution of the spatially-averaged, normalized gradient energy density in terms of the number of $e$-folds for seven BPs outlined in Table~\ref{tab:BPs}. As before, different colors indicate different values of $\ln[\xi^2/(4\alpha\lambda)]$, and the vertical dashed lines depict the points where the oscillation-averaged gradient energy density reaches its maximum. We observe that as the nonminimal coupling parameter $\xi$ increases, the magnitude of the gradient energy density becomes larger, and the time it takes for the gradient energy density to reach its maximum becomes shorter, implying more efficient preheating.}
    \label{fig:grad}
\end{figure}
The evolution of the spatially-averaged, normalized gradient energy density for all BPs is presented in Fig.~\ref{fig:grad}. Different colors again indicate different values of $\ln[\xi^2/(4\alpha\lambda)]$. As in Fig.~\ref{fig:energydensity-1-and-7}, the vertical dashed lines depict the points where the gradient energy density averaged over oscillations reaches its maximum. One can easily notice that as we move from the $R^2$-like parameter region to the Higgs-like parameter region, preheating becomes more efficient. In other words, as the nonminimal coupling parameter $\xi$ increases, the magnitude of the gradient energy density becomes larger, and the time it takes for the gradient energy density to reach its maximum becomes shorter. The same behavior is found in Ref.~\cite{Bezrukov:2020txg}.

The exponential enhancement of the gradient energy density hints at possible generations of GWs \cite{Dufaux:2007pt,Amin:2014eta,Lozanov:2020zmy}. In the next subsection, based on the results of lattice simulation, we discuss the GW signatures of preheating in the Higgs--$R^2$ model.

\subsection{Gravitational Waves}
\label{subsec:GWs}
We are now in a position to estimate GWs sourced by the enhancement of inhomogeneities during the preheating phase. For a comprehensive review on GWs, readers may refer to Refs.~\cite{Maggiore:2007ulw,Maggiore:2018sht}. We consider a transverse-traceless part of the tensor perturbation on the FRW metric,
\begin{align}
    ds^2 = a^2\left[
    -d\tau^2 + \left(
    \delta_{ij} + h_{ij}
    \right)dx^idx^j
    \right]
    \,.
\end{align}
Working in Fourier space, and introducing a new variable $\bar{h}_{ij} \equiv ah_{ij}$, we obtain the GW equation as
\begin{align}
    \bar{h}_{ij}'' + \left(
    k^2 - \frac{a''}{a}\right)\bar{h}_{ij} = 2a\frac{T^{\rm TT}_{ij}}{M_{\rm P}^2}
    \,,\label{eqn:GWeqn}
\end{align}
where the source term $T^{\rm TT}_{ij}$ is the transverse-traceless part of the stress-energy tensor $T_{ij}$. The transverse-traceless part can be extracted by utilizing the projection operator as follows:
\begin{align}
    T^{\rm TT}_{ij} =
    \sum_{m,n}\left(
    P_{im}P_{jn} - \frac{1}{2}P_{ij}P_{mn}
    \right)T_{mn}
    \,,
\end{align}
where
\begin{align}
    P_{ij} = \delta_{ij} - \frac{k_i k_j}{k^2}
    \,.
\end{align}
For a two-field model with a kinetic coupling function as our model under consideration \eqref{eqn:einstein-action}, the transverse-traceless part of the stress-energy tensor is, up to the first order in the tensor perturbation, given by
\begin{align}
    T_{ij}^{\rm TT} = \left[\partial_i\chi\partial_j\chi + f(\chi)\partial_i\phi\partial_j\phi\right]^{\rm TT}
    \,.
\end{align}
\begin{figure*}[ht!]
    \centering
    \includegraphics[width=.48\textwidth]{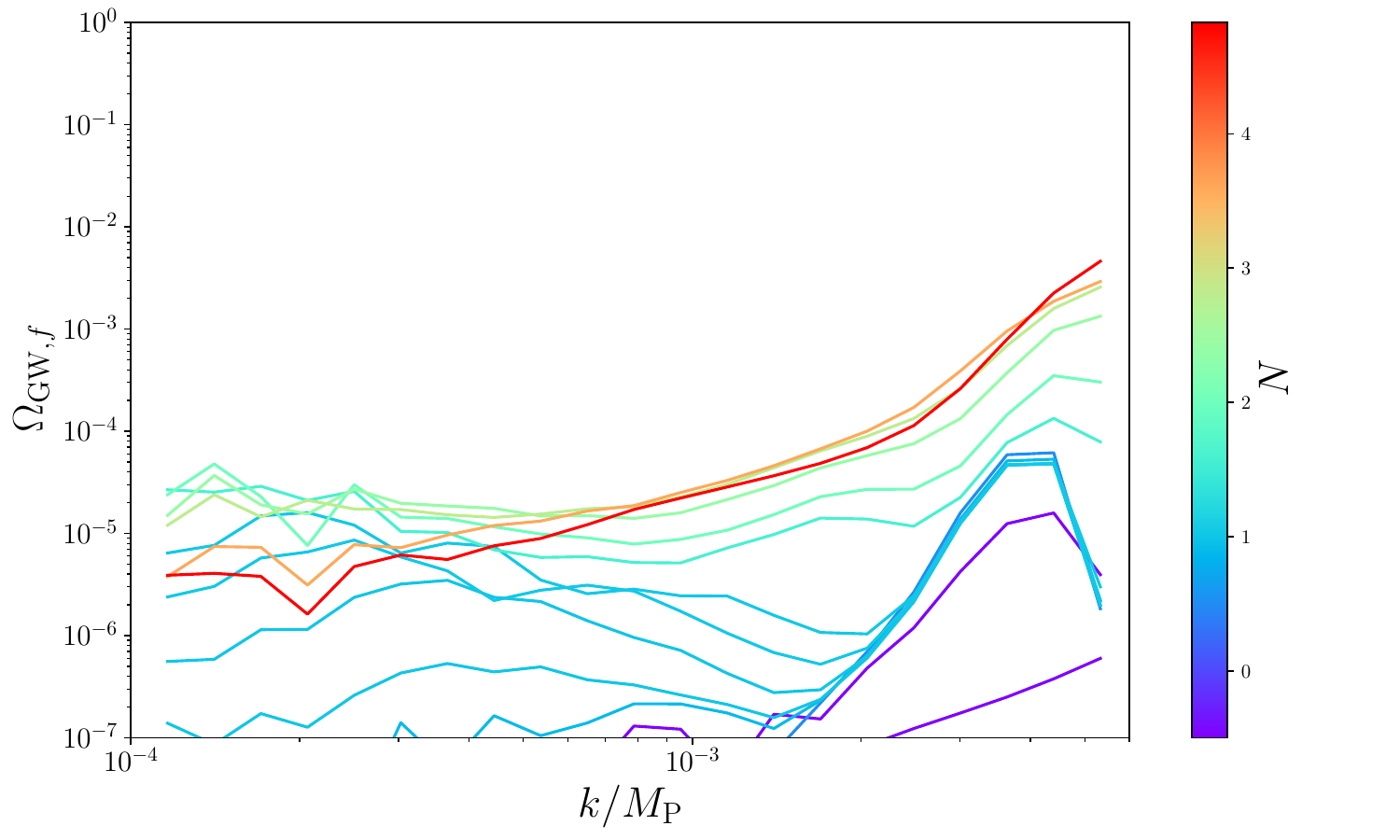}\;
    \includegraphics[width=.48\textwidth]{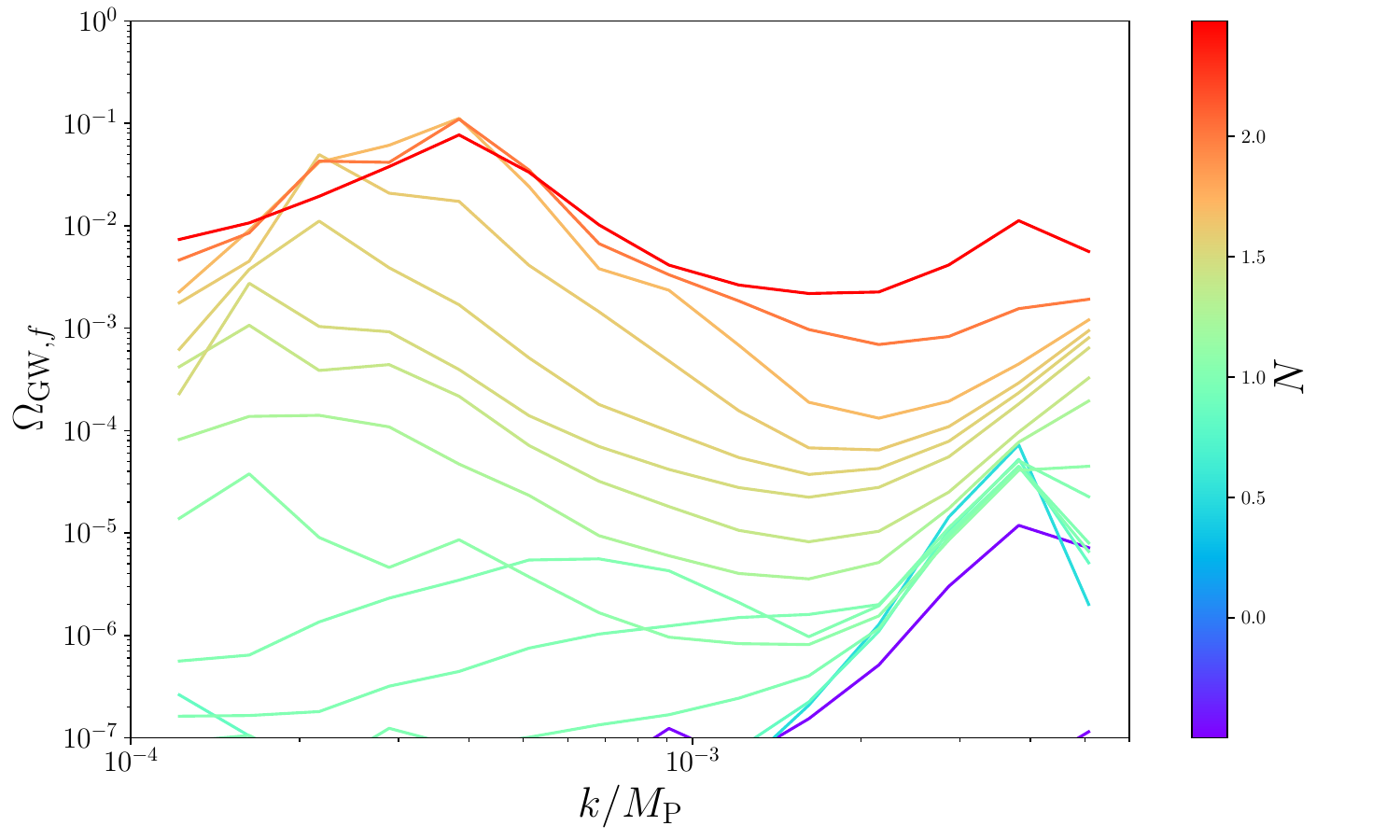}
    \caption{Evolution of the GW spectrum for BP1 (left) and BP7 (right) outlined in Table~\ref{tab:BPs}. The $x$-axis represents the wavenumber $k$ normalized by the reduced Planck mass $M_{\rm P}$, and the color depicts the number of $e$-folds during the simulation. The $y$-axis depicts the GW spectrum at the time of generation, $\Omega_{{\rm GW},f}$. The growth and peak behavior in the high-$k$ region is an artifact coming from our choice of the momentum cutoff of $10^3H_*$, where $H_*$ is the inflationary Hubble scale; see also Fig.~\ref{fig:gws-set4-cutoff}. We observe the growth of the GW spectrum in accordance with the enhancement of the inhomogeneities in the fields observed in Fig.~\ref{fig:energydensity-1-and-7}.}
    \label{fig:gws-bp1-bp7}
\end{figure*}

GW signatures arising from preheating have been extensively investigated in, for example, Ref.~\cite{Dufaux:2007pt}. Furthermore, in Refs.~\cite{Krajewski:2022ezo,Adshead:2024ykw}, the GW spectrum is discussed for a two-field model with a kinetic coupling function. In the current work, following the method developed in Ref.~\cite{Krajewski:2022ezo}, which adopts the second-order Magnus expansion \cite{BLANES2012875}, we compute the GW spectrum by directly solving the GW equation \eqref{eqn:GWeqn}. For completeness, we briefly discuss the method; see also Appendix A of Ref.~\cite{Krajewski:2022ezo}. The GW equation \eqref{eqn:GWeqn} can be re-written in the compact form as follows:
\begin{align}
    Y' = MY + F\,,
\end{align}
where
\begin{align}
    Y &\equiv 
    \begin{pmatrix}
        \bar{h}_{ij} \\ 
        \bar{h}'_{ij}
    \end{pmatrix}
    \,,\\
    M &\equiv
    \begin{pmatrix}
        0 & 1 \\
        a''/a - k^2 & 0
    \end{pmatrix}
    \,,\\
    F &\equiv
    \begin{pmatrix}
        0 \\
        2aT^{\rm TT}_{ij}/M_{\rm P}^2
    \end{pmatrix}
    \,.
    \label{eqn:YMF-definitions}
\end{align}
The solution is then formally given by \cite{BLANES2012875}
\begin{align}
    Y(\tau+d\tau) &= \Phi(\tau+d\tau,\tau)Y(\tau) 
    \nonumber\\&\qquad
    + \int_{\tau}^{\tau+d\tau}d\tau' \, \Phi(\tau+d\tau,\tau')F(\tau')
    \,,\label{eqn:Yeqn}
\end{align}
where $\Phi$ is the solution of the homogeneous equation,
\begin{align}
    \Phi' = M\Phi\,,
\end{align}
whose solution is given, using the Magnus approximation, by
\begin{align}
    \Phi(\tau+d\tau,\tau) = \exp\left[
    \int_{\tau}^{\tau+d\tau} d\tau' \, M(\tau') + \mathcal{O}\left(d\tau^4\right)
    \right]
    \,.
\end{align}
We may approximate the integrals using the trapezoidal rule and obtain
\begin{align}
    \Phi(\tau+d\tau,\tau) \approx 
    \exp\left\{
    \frac{d\tau}{2}\left[
    M(\tau) + M(\tau+d\tau)
    \right]
    \right\}\,,
\end{align}
and upon inserting it in Eq.~\eqref{eqn:Yeqn}, we have
\begin{align}
    Y(\tau+d\tau) &\approx 
    \left[
    Y(\tau) + \frac{d\tau}{2}F(\tau)
    \right]
    \nonumber\\&\quad\times
    \exp\left\{
    \frac{d\tau}{2}\left[
    M(\tau) + M(\tau+d\tau)
    \right]
    \right\}
    \nonumber\\&\quad
    +\frac{d\tau}{2}F(\tau+d\tau)
    \,.\label{eqn:Ysol-continuous}
\end{align}
For numerical calculations, it is more convenient to express the solution \eqref{eqn:Ysol-continuous} in a discretized manner with the time step $n$ as
\begin{align}
    Y_{n+1} &= 
    \left(
    Y_n + \frac{d\tau}{2}F_n
    \right)\exp\left\{
    \frac{d\tau}{2}\left(
    M_n + M_{n+1}
    \right)
    \right\}
    \nonumber\\&\qquad
    +\frac{d\tau}{2}F_{n+1}
    \,.\label{eqn:Ysol-discrete}
\end{align}
Finally, using the definitions of $Y$, $M$, and $F$ given in Eq.~\eqref{eqn:YMF-definitions}, we can express the solutions for $\bar{h}$ and $\bar{h}'$ explicitly as follows:
\begin{align}
    \bar{h}_{ij,n+1} &=
    \cos(\omega d\tau)\bar{h}_{ij,n}
    \nonumber\\&\quad
    +\frac{1}{\omega}\sin(\omega d\tau)\left(
    \bar{h}'_{ij,n}
    +a_nd\tau\frac{T^{\rm TT}_{ij,n}}{M_{\rm P}^2}
    \right)
    \,,\label{eqn:hij}\\
    \bar{h}'_{ij,n+1} &=
    \cos(\omega d\tau)\left(
    \bar{h}'_{ij,n}
    +a_nd\tau\frac{T^{\rm TT}_{ij,n}}{M_{\rm P}^2}
    \right)
    \nonumber\\&\quad
    -\omega\sin(\omega d\tau)\bar{h}_{ij,n} 
    +a_{n+1}d\tau\frac{T^{\rm TT}_{ij,n+1}}{M_{\rm P}^2}
    \,,\label{eqn:hijprime}
\end{align}
where
\begin{align}
    \omega^2 \equiv k^2 - \frac{a''_{n+1}}{2a_{n+1}} - \frac{a''_n}{2a_n}\,.
\end{align}

The physical quantity in which we are most interested is the GW spectrum, which is defined as
\begin{align}
    \Omega_{\rm GW} =
    \frac{1}{\rho_c}
    \frac{d\rho_{\rm GW}}{d\ln k}
    \,,\label{eqn:OmegaGWh2}
\end{align}
where $\rho_c \equiv 3M_{\rm P}^2H^2$ is the critical energy density, and $\rho_{\rm GW}$ is the GW energy density given by
\begin{align}
    \rho_{\rm GW} \equiv T^{\rm GW}_{00} = \frac{M_{\rm P}^2}{4} \sum_{i,j}\left\langle
    \dot{h}_{ij}(t,{\bf x})
    \dot{h}_{ij}(t,{\bf x})
    \right\rangle\,.
\end{align}
Here, $T^{\rm GW}_{00}$ is the 00-component of the Isaacson stress-energy tensor \cite{Isaacson:1968hbi,Isaacson:1968zza}, and $\langle\cdots\rangle$ denotes an average over a comoving volume $V_3$. Moving to Fourier space and taking the subhorizon limit, the GW energy density can be expressed in terms of $\bar{h}$ as \cite{Dufaux:2007pt}
\begin{align}
    \rho_{\rm GW} \approx 
    \frac{M_{\rm P}^2}{4a^4V_3}\sum_{i,j}\int \frac{d^3k}{(2\pi)^3}\bar{h}'_{ij}(\tau,{\bf k})\bar{h}'^*_{ij}(\tau,{\bf k})
    \,.\label{eqn:rhoGW}
\end{align}
From the results of lattice simulations, we calculate the stress-energy tensor $T^{\rm TT}_{ij}$. We then utilize Eqs.~\eqref{eqn:hij} and \eqref{eqn:hijprime} to find solutions for the GW tensors $\bar{h}_{ij}$ and $\bar{h}'_{ij}$. Substituting the solutions to Eq.~\eqref{eqn:rhoGW}, we obtain the GW energy density. 

Evolution of the GW spectrum $\Omega_{\rm GW}$ \eqref{eqn:OmegaGWh2} is presented in Fig.~\ref{fig:gws-bp1-bp7} for two BPs, namely BP1 (left panel) and BP7 (right panel), outlined in Table~\ref{tab:BPs}. The color depicts the number of $e$-folds at the time of evaluating the GW spectrum. In other words, we compute the GW spectrum as if the GWs are emitted at a certain $e$-folds or, equivalently, a certain $\tau_f$. The subscript $f$ in $\Omega_{{\rm GW},f}$ makes this point apparent. The $x$-axis represents the wavenumber $k$ normalized by the reduced Planck mass $M_{\rm P}$. The growth and peak behavior in the high-$k$ region is an artifact coming from our choice of the momentum cutoff of $10^3H_*$, where $H_*=M_{\rm P}/\sqrt{12(4\alpha+\xi^2/\lambda)}$ is the inflationary Hubble scale; see also Fig.~\ref{fig:gws-set4-cutoff}. We observe that the GW spectrum grows in accordance with the enhancement of the inhomogeneities in the fields observed in Fig.~\ref{fig:energydensity-1-and-7} as well as in Fig.~\ref{fig:grad}.

In order to compute the GW spectrum at present, a proper redshift factor should be accounted for to the GW spectrum obtained from the lattice simulations. Following Ref.~\cite{Dufaux:2007pt}, the present-day GW spectrum is related to the GW spectrum at the time of formation as
\begin{align}
    \Omega_{\rm GW}h^2 \approx 
    9.3 \times 10^{-6}\frac{1}{\rho_{c,f}}\frac{d\rho_{\rm GW}}{d\ln k}\bigg\vert_{\tau=\tau_f}
    \,,\label{eqn:GWspectrum-today}
\end{align}
where $h \approx 0.67$ is the scaling factor for the Hubble parameter, $\rho_{c,f}$ is the critical energy density at $\tau=\tau_f$, and we choose $\tau_f$ to be the time at which the oscillation-averaged gradient energy density reaches its maximum, denoted by the vertical dashed lines in Fig.~\ref{fig:grad}.
Moreover, the wavenumber $k$ is related to the frequency today through \cite{Dufaux:2007pt}
\begin{align}
    f = 4 \times 10^{10}
    \left(
    \frac{k}{a_f\rho_{c,f}^{1/4}}
    \right)
    \, {\rm Hz}
    \,,\label{eqn:freq-k}
\end{align}
with $a_f$ being the scale factor at $\tau=\tau_f$.
It is worth stressing that in arriving at Eqs.~\eqref{eqn:GWspectrum-today} and \eqref{eqn:freq-k}, we have assumed that the scale factor at the time of thermal equilibrium establishment takes roughly the same value as the scale factor at $\tau=\tau_f$~\cite{Dufaux:2007pt}.

\begin{figure}[t!]
    \centering
    \includegraphics[width=0.98\linewidth]{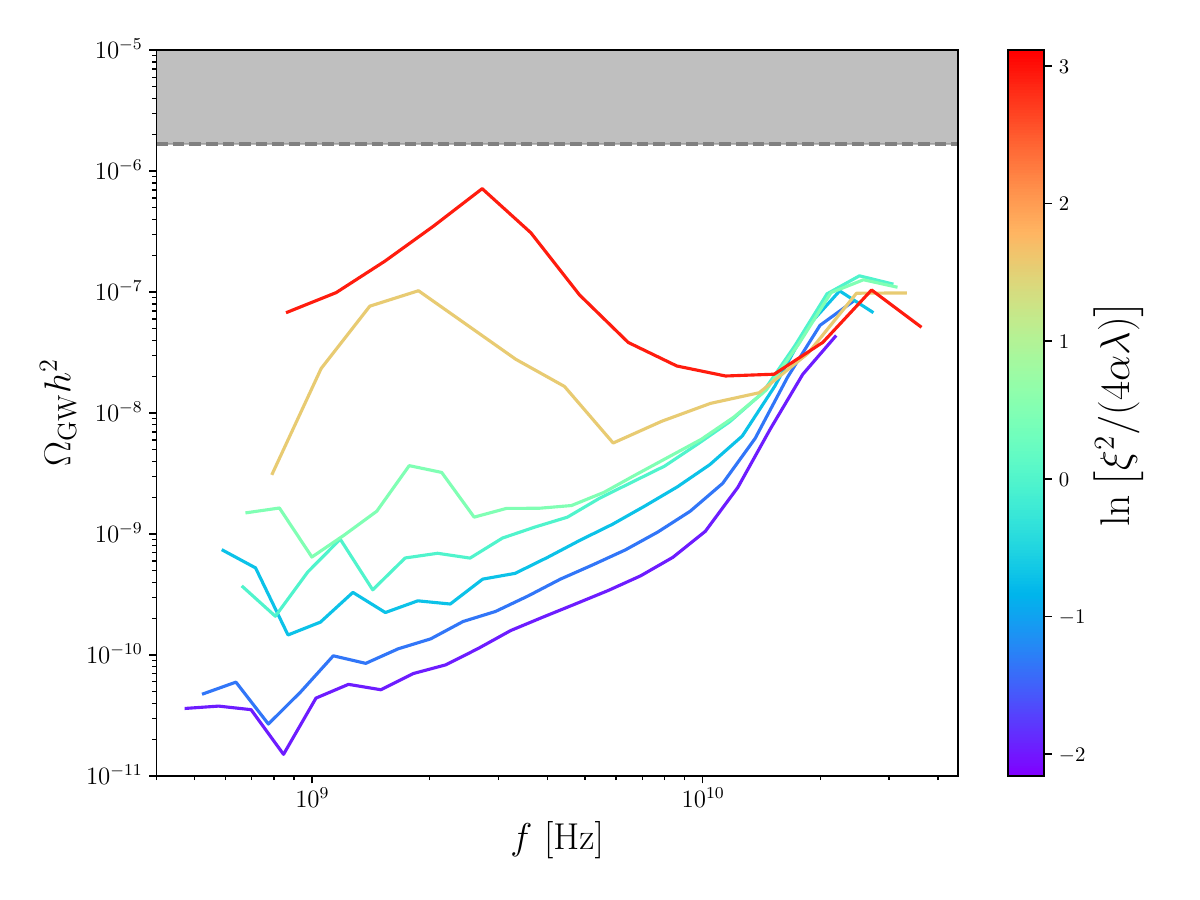}
    \caption{Present-day GW spectrum $\Omega_{\rm GW}h^2$ for all the BPs outlined in Table~\ref{tab:BPs} in terms of the present-day frequency $f$. The spectrum at generation is evaluated at the time where the oscillation-averaged gradient energy density reaches its maximum denoted by the vertical dashed lines in Fig.~\ref{fig:grad}. Different colors indicate different values of $\ln[\xi^2/(4\alpha\lambda)]$. The gray region, $\Omega_{\rm GW}h^2 \gtrsim 1.68 \times 10^{-6}$, indicates a bound coming from the effective relativistic degrees of freedom based on the CMB observation. We see that a stronger enhancement in inhomogeneities in the fields leads to a higher GW spectrum.}
    \label{fig:gws-max}
\end{figure}
In Fig.~\ref{fig:gws-max}, the present-day GW spectrum \eqref{eqn:GWspectrum-today} for all the seven BPs considered in our analysis is presented. The GW spectrum at $\tau=\tau_f$ is evaluated at the time where the gradient energy density reaches its maximum, which is denoted by the vertical dashed lines in Fig.~\ref{fig:grad}. As before, different colors indicate different values of $\ln[\xi^2/(4\alpha\lambda)]$, which is an indicator for whether the inflationary scenario is Higgs-like or $R^2$-like.
The gray region comes from a bound on the GW spectrum based on the CMB observation. GWs may contribute to the effective relativistic degrees of freedom $N_{\rm eff}$, which, in turn, can affect the physics of nucleosynthesis. The energy density of GWs is related to the effective relativistic degrees of freedom via $\Omega_{\rm GW}h^2 \simeq 5.61\times 10^{-6}(N_{\rm eff}-N_{\rm eff}^{\rm SM})$~\cite{Maggiore:1999vm,Maggiore:2018sht} (see also Ref.~\cite{Caprini:2018mtu}), where $N_{\rm eff}^{\rm SM}\simeq 3.046$ is the Standard Model value. Taking $N_{\rm eff}-N_{\rm eff}^{\rm SM} < 0.3$ (95\% C.L.) \cite{Planck:2018vyg} gives the bound of $\Omega_{\rm GW}h^2 \lesssim 1.68 \times 10^{-6}$.
We see that a larger value of $\xi$ brings a greater increase in the GW spectrum. It aligns with our expectation that a stronger enhancement in inhomogeneities in the fields leads to a higher GW spectrum.
The prominent peak signal is observed at the frequency $f \simeq 2\times 10^9$ Hz.
We note again that the high-frequency peak at $f\simeq (2-3)\times10^{10}$ Hz in the GW spectrum, which corresponds to the momentum cutoff, $k \simeq 10^3 H_*$, is spurious; varying the cutoff changes the position of the high-frequency peak. This feature is explicitly demonstrated for BP4 in Fig.~\ref{fig:gws-set4-cutoff}; three different values of the momentum cutoff, $\Lambda = 700$ (dashed), 1000 (solid), and 1300 (dot-dashed), are chosen with the fixed lattice size of $64^3$. We observe that the spurious peak moves towards the higher-frequency region as the cutoff increases, indicating that such a peak is an artifact.

\begin{figure}[t!]
    \centering
    \includegraphics[width=0.98\linewidth]{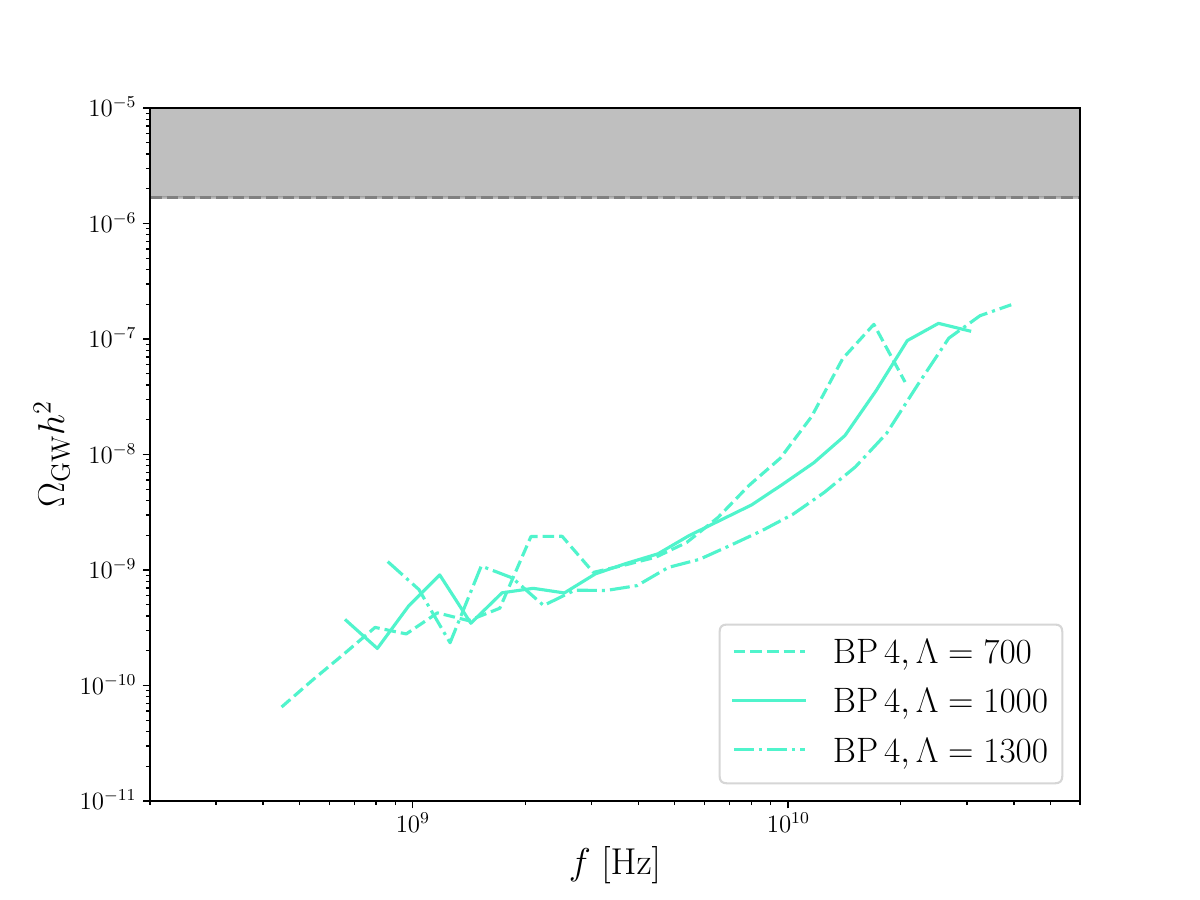}
    \caption{Present-day GW spectrum $\Omega_{\rm GW}h^2$ for BP4 in terms of the present-day frequency $f$ for three different choices of the momentum cutoff $\Lambda=700$ (dashed), 1000 (solid), and 1300 (dot-dashed). The lattice size is chosen to be $64^3$ for all three cases. As in Fig.~\ref{fig:gws-max}, the gray region, $\Omega_{\rm GW}h^2 \gtrsim 1.68 \times 10^{-6}$, indicates a bound coming from the effective relativistic degrees of freedom based on the CMB observation. We observe that the peak at the very high frequency, $f \simeq (2-3)\times 10^{10}$ Hz, depends on the choice of the momentum cutoff; as the cutoff increases, the spurious peak moves towards the higher-frequency region.}
    \label{fig:gws-set4-cutoff}
\end{figure}

\section{Conclusion}
\label{sec:conclusion}
In this work, we have conducted a comprehensive investigation of the preheating dynamics and the associated gravitational wave signatures in the Higgs--$R^2$ inflationary model. Our analysis covers a wide parameter space from the Higgs-like scenario to the $R^2$-like scenario, providing a thorough understanding of the post-inflationary behavior of the model. Performing dedicated, detailed lattice simulations, we have tracked the evolution of the fields as well as various energy density components, with a particular focus on the inhomogeneous modes captured by the gradient energy density.

A characteristic dependence of preheating behaviors on the model parameters has been observed. We have shown that as the nonminimal coupling parameter $\xi$ increases, the preheating process becomes more efficient. This tendency is manifested in two ways, first through larger amplitudes of the gradient energy density, and second through shorter duration for the gradient energy density to reach its maximum value. Such an observation can be understood as a consequence of the stronger tachyonic instability in the Higgs-like regime. We further noted that the result of numerical lattice simulations is in good agreement with the analytical study of the field perturbations.

Based on the outcome of the lattice simulations, we have numerically computed gravitational waves sourced by the inhomogeneities. Our computation has revealed a direct correlation between the efficiency of preheating and the amplitude of the gravitational wave spectrum. Aligning with the behavior of the gradient energy density, larger values of the nonminimal coupling parameter $\xi$ lead to more pronounced gravitational wave production. We have shown that the peak amplitude of the gravitational wave spectrum $\Omega_{\rm GW}h^2$ varies by several orders of magnitude across our seven benchmark points.

The significant variation in the gravitational wave spectrum provides a potential way to distinguish between different parameter regimes of the Higgs--$R^2$ model. However, the gravitational wave frequencies lie in the range of $10^{9-10}$ Hz, which is beyond the reach of current gravitational wave experiments. We believe that our study further advocates the importance and the necessity of high-frequency gravitational wave detectors.

Throughout the analysis, we have set the vacuum expectation value of the $\phi$ field, namely the Higgs field, to zero, $v=0$. A more realistic picture would have a nonzero $v$. It is known that the inclusion of a nonzero $v$ may lead to the production of oscillons in the absence of the $R^2$ term. Furthermore, introducing additional field contents, which is necessary for a realistic particle physics-based Higgs field, can modify the preheating dynamics, thereby predicting different gravitational wave spectrum. Finally, we note that for the model we considered in the current work, a different formulation of general relativity, such as the Palatini formulation \cite{Palatini:1919ffw,Ferraris:1982wci}, could be used. We plan to return to these issues in the future.

\acknowledgments
J.K., Z.Y., and Y.Z. are supported by National Natural Science Foundation of China (NSFC) under Grant No. 12475060 and by Shanghai Pujiang Program 24PJA134.
Y.Z. is also supported by the Fundamental Research Funds for the Central Universities, the Project 12047503 supported by NSFC, and Project 24ZR1472400 sponsored by Natural Science Foundation of Shanghai.



\bibliographystyle{apsrev4-2}
\bibliography{main}

\end{document}